\documentclass[12pt,final,5p,times,twocolumn,]{elsarticle}

\usepackage{amssymb}
\usepackage[fleqn]{amsmath}
\usepackage{color}
\usepackage{bm}
\usepackage{booktabs}
\usepackage{caption}
\newcommand{\tabincell}[2]{\begin{tabular}{@{}#1@{}}#2\end{tabular}}
\usepackage{stfloats}
\biboptions{numbers,sort&compress}

\begin{document}
\captionsetup[figure]{labelfont={bf},labelformat={default},labelsep=period,name={Fig.}}
\captionsetup[table]{labelfont={bf},labelsep=space}

\begin{frontmatter}


	
\title{Designing thermal energy harvesting devices with natural materials through optimized microstructures}


%\author{Qingxiang Ji$^{a,b}$, Xueyan Chen$^{a,b}$, Jun Liang$^{c}$, Vincent Laude$^{b}$, S\'ebastien Guenneau$^{d}$, Guodong Fang$^{a,*}$ and Muamer Kadic$^{b}$}
\author{Qingxiang Ji$^{a,b}$}
\author{Xueyan Chen$^{a,b}$}
\author{Jun Liang$^{c}$}
\author{Vincent Laude$^{b}$}
\author{S\'ebastien Guenneau$^{d}$}
\author{Guodong Fang $^{a,}$ \corref{mycorrespondingauthor}}
\cortext[mycorrespondingauthor]{Corresponding author\\ E-mail address: fanggd@hit.edu.cn}
%\ead{fanggd@hit.edu.cn}
\author{Muamer Kadic$^{b}$}

\address{%
		$^{a}$ \quad National Key Laboratory of Science and Technology on Advanced Composites in Special Environments, Harbin Institute of Technology, Harbin, 150001, China\\
		$^{b}$ \quad Institute FEMTO-ST, CNRS, University Bourgogne Franche-Comt{\'e}, 25000 Besan\c{c}on, France\\
		$^{c}$ \quad Institute of Advanced Structure Technology, Beijing Institute of Technology, Beijing 100081, China\\
		$^{d}$ \quad UMI 2004 Abraham de Moivre-CNRS, Imperial College, London SW7 2AZ, UK}

%\fntext[myfootnote]{fanggd@hit.edu.cn}

\begin{abstract}
Metamaterial thermal energy devices obtained from transformation optics have recently attracted wide attention due to their vast potential in energy storage, thermal harvesting or heat manipulation. However, these devices usually require inhomogeneous and extreme material parameters which are difficult to realize in large-scale applications. Here, we demonstrate a general process to design thermal harvesting devices with available natural materials through optimized composite microstructures. We apply two-scale homogenization theory to obtain effective properties of the microstructures. Optimal Latin hypercube technique, combined with a genetic algorithm, is then implemented on the microstructures to achieve optimized design parameters. The optimized microstructures can accurately approximate the behavior of transformed materials. We design such devices and numerically characterize good thermal-energy harvesting performances. To validate the wide-range application of our approach, we illustrate other types of microstructures that mimic well the constitutive parameters. The approach we propose can be used to design novel thermal harvesting devices available with existing technology, and can also act as a beneficial vehicle to explore other transformation optics enabled designs.

\end{abstract}


\end{frontmatter}


\section{Introduction}
Transformation optics was first proposed to perform cloaking on electromagnetic waves, based on form-invariance of the governing equations after coordinates transformation \cite{pendry2006controlling}. Since then the concept was promoted in various physical fields \cite{kadic2013metamaterials,milton2006cloaking,buckmann2014elasto,achaoui2020cloaking,kadic2020elastodynamic,schittny2014invisibility,schurig2006metamaterial,cummer2007one,guenneau2012transformation}, creating miscellaneous devices, like invisible cloaks \cite{farhat2009ultrabroadband}, carpets \cite{li2008hiding}, invisible sensors \cite{yang2015invisible}, illusion devices \cite{liu2012dc,hu2018illusion} or hyperlenses \cite{liu2007far,li2009experimental,kadic2011plasmonic}. 

Over the past decades, the technique of thermal energy harvesting and management,  which can collect and store heat energy from the ambient surroundings, has attracted renewed interest \cite{freeman2017small,amin2017thermal}. Transformation thermodynamics, a counterpart of transformation-optics, has been proposed to guide heat flux in thermal management and generate novel thermal meta-devices, such as heat cloaks, thermal energy harvesting devices, thermal sensors, etc \cite{fan2008shaped,zeng2014experimental,schittny2013experiments,han2014full,peng2019three,huang2019macroscale}. Thermal energy harvesting devices, which can focus and harvest ambient thermal energy without severe perturbations to the heat profile outside the devices, have vast potential in improving the energy-conversion efficiency of existing technologies \cite{xu2018achieving}. 

A challenge for thermal meta-devices is that they often require inhomogeneous and anisotropic constitutive parameters which are difficult to realize especially for large-scale applications. As the heterogenous constitutive profile is position-dependent and continuous, some form of discretization is required. Moreover, anisotropic materials can be approximated by structures of thin and alternating layers based on effective medium theory \cite{schurig2006metamaterial,chen2010transformation}. Thermal devices following this principle were fabricated and experimentally characterized \cite{schittny2013experiments,narayana2012heat}. In these works, step-wise approximation of the ideal parameters were made, which sacrificed the performances to reduce fabrication difficulty.
Other researchers realized thermal cloaks with bulk isotropic materials using the scattering-cancellation approach \cite{han2014experimental,zhang2020ellipsoidal,xu2014ultrathin,han2018full}, which was then extended to the design of thermal concentrating devices. 
Following this approach, a class of solar-shaped thermal harvesting devices were demonstrated, which could manipulate and concentrate the heat flux using natural materials without singularities \cite{chen2015experimental,han2013theoretical}.
However, these meta-devices are always shaped as regular profiles (cylinder, sphere or ellipse), as the scattering-cancellation method is non-trivial for irregular or complex shapes. 

Except for the mentioned strategies, do we have methods that can both offer good performance and fabrication simplicity? The answer is positive. Researchers \cite{mei2007effective,torrent2008acoustic} demonstrated that the constitutive medium can be approximated by microstructures in the context of acoustic metamaterials. Thanks to similarities between the governing equations for acoustic and thermal fields, a similar approach can be implemented for thermal metamaterials. Then the key problem is to determine the effective property of the built microstructure and find the proper parameters of the microstructure to mimic desired transformed parameters. Ji \textit{et al.} \cite{ji2018achieving} achieved thermal concentration using fiber reinforced microstructures based on simplified effective medium theory. Pomot \textit{et al.} \cite{pomot2020acoustic} realized acoustic cloaking by microstructures with several types of perforations, combined with a genetic algorithm. We note that the partial differential equations (PDEs) therein have the same structure to the heat equation in the static limit, in which case one simply has to replace the density by the conductivity and all the asymptotic analysis carries through. Moreover, PDEs in Ref. \cite{pomot2020acoustic} are supplied with homogeneous Neumann boundary conditions that stand for rigid walls in acoustics, and insulating walls in thermodynamics. It is thus tempting to implement homogenization and effective medium theories similar to those already used in acoustic metamaterials, however bearing in mind that in the dynamic regime the acoustic wave (elliptic) PDE and the heat diffusion (parabolic) PDE are very different in nature.

In our work we focus on the thermal field and establish a general design road map to obtain realizable thermal harvesting devices utilizing micro-structures. We apply two-scale homogenization theory to determine the equivalent thermal properties and employ the Optimal Latin-Hypercube technique to obtain desired design parameters. Beyond this problem, the process is applicable to other microstructures and to wave problems (such as in acoustics). We stress that because of their complexity, the microstructures we investigate here would be challenging to achieve otherwise. Finally we numerically characterize such thermal harvesting devices with natural materials and verify their harvesting efficiency by finite element simulations.

\section{Methods}
\subsection{Concept design}
We recall the heat conduction equation without heat sources
\begin{equation} \label{GrindEQ__2.6_}
\centering
\nabla \cdot (k \nabla T)-\rho c \frac{\partial T}{\partial t}=0,
\end{equation}
where $\nabla$ is the gradient, $T$ is the temperature, $\rho c$ is the product of density by heat capacity and $k$ denotes the heat conductivity. Following the theory of transformation optics \cite{pendry2006controlling} and thermodynamics \cite{guenneau2012transformation}, the governing equation will remain unchanged under a coordinate transformation when the transformed parameters satisfy
\begin{equation} \label{GrindEQ__2.1_}
\centering
k' =\frac{JkJ^{T} }{\det (J)} \ \ \rm{and} \ \ (\rho c)'=\frac{\rho c}{\det (J)},
\end{equation}
where $J$ is the Jacobian of the geometric transformation, $J^T$ its transpose and det($J$) its determinant. We first consider steady-state case where parameters $\rho c$ vanish.
The transformed conductivity $k'$ derived by Eq. \eqref{GrindEQ__2.1_} is usually anisotropic and space dependent, which is difficult to achieve in practice. To remove the singularities and make the proposed device simpler to realize, we choose the following nonlinear transformation that maps domain $\Omega(r)$ onto domain $\Omega'(r')$: 
\begin{equation} \label{GrindEQ__2.5_}
\centering
r' = r_2 \; \left( \frac{r}{r_2} \right)^C,
\end{equation}
\noindent which yields constant relative radial ($k_r'$) and tangential heat conductivity ($k_\theta'$) in cylindrical coordinates as
\begin{equation} \label{GrindEQ__2.2_}
\centering
k_r' =C \ \ \rm{and} \ \ k_\theta'=\frac{1}{C},
\end{equation}
\noindent where C is a constant with $C>1$. Here, $r_1$ and $r_2$ are the inner radius and the outer radius of the designed shell region, respectively (see details in the supplemental material).

%Ref. \cite{han2013theoretical} demonstrated thermal cells concentrating heat nearly perfectly with constant anisotropic conductivity which was much simplified. 
In stark contrast to the material parameters obtained from rigorous transformation optics, such a device is homogeneous in materials composition and its performance is only determined by the thermal-conduction anisotropy (characterized by constant $C$; see details in the supplemental material). We employ this geometric transformation as it avoids the need for extreme spatially-varying parameters and thus the transformed medium will be much easier to implement in practice. Moreover, it is enough for the proof-of-concept demonstration. We stress that the method we describe is indeed also applicable to the design of rigorous transformation-based thermal devices. 
 
\begin{figure}[!tbh]
	\centering
	\includegraphics*[width=9cm, keepaspectratio=true]{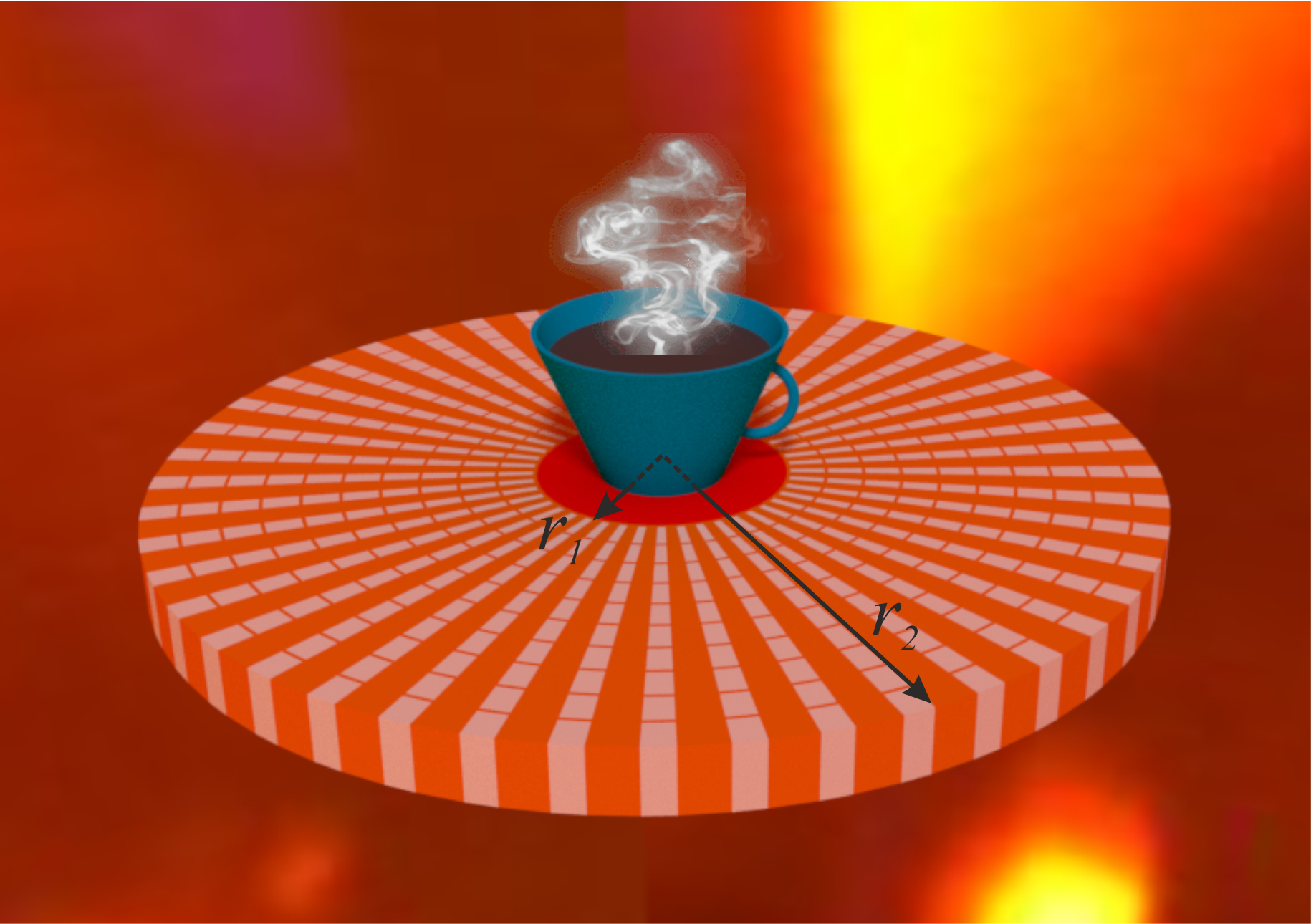}
	\caption{Schematic representation of a possible realization of the thermal harvesting device. The shell displays some periodicity along the azimuthal direction, which is a hallmark of a concentrator. Such a device can harvest thermal energy from its surroundings. Heat flows are concentrated into the inner domain $r'< r_1$ thanks to the designed shell region $r_1 \leq r'\leq r_2$. }
	\label{fig-illustration}
\end{figure}

For clarity, we outline the design process for the general case with a specific example. Assume that we create a cylindrically symmetric thermal harvesting device with inner radius $ r_1=0.01 \rm{~m}$ and outer radius $ r_2=0.04 \rm{~m}$ (see Fig. \ref{fig-illustration}). The proposed device can harvest thermal energy from the surroundings and concentrate it into the inner domain $r'<r_1$. Heat energy density in the inner domain is thus significantly increased. 
In our design, the thermal conductivity of the background is $k_b=132 \rm{~W m^{-1} K^{-1}}$. We set $C=2$, which implies that for the shell region we have $ k_r=264 \rm{~W m^{-1} K^{-1}}$, $k_{\theta}=66 \rm{~W m^{-1} K^{-1}}$. Now we turn to the realization of shell region parameters by two natural materials $A$ and $B$ through optimized microstructures. 

\subsection{Homogenization of the heat conduction equation}

\begin{figure}[!tbh]
	\centering
	\includegraphics*[width=9cm, keepaspectratio=true]{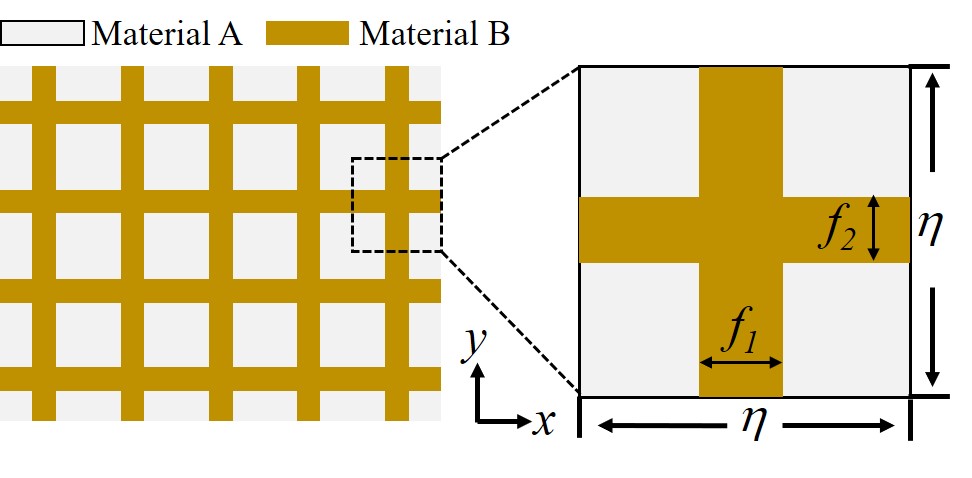}
	\caption{Schematic illustration of 2D periodic lattices defined by 2 parameters (area fractions $f_1$ and $f_2$).  We choose this microstructure to mimic the desired medium of the shell region through homogenization and optimization.	}
	\label{fig-cell}
\end{figure}

The material distributions in Eq. \eqref{GrindEQ__2.2_} need to be mapped onto a microstructure exhibiting prescribed constitutive parameters. We build a medium with identical elementary cells repeating periodically in space (see Fig. \ref{fig-cell} ). Generally, the well-established effective medium theory plays a dominant role in determining effective properties and is easy and direct for simple geometries. When designing meta-devices with complex-shaped structures, however, it would be far from trivial to evaluate the equivalent properties. Here we apply instead two-scale homogenization theory \cite{zolla2003artificial} to determine effective parameters with an asymptotic approach. 

\begin{figure}
	\centering
	\includegraphics*[width=9cm, keepaspectratio=true]{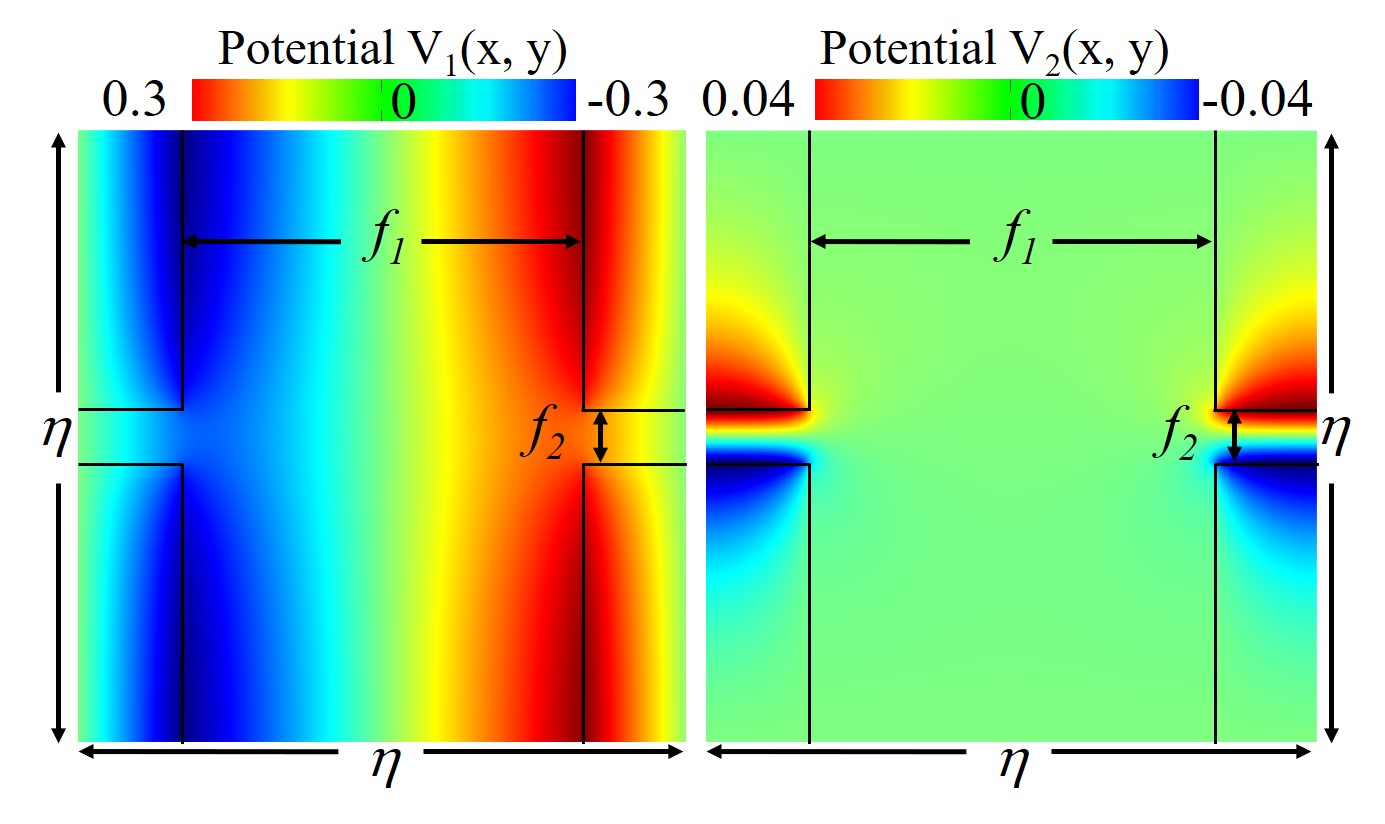}
	\caption{Potentials $V_1(x,y)$ and $V_2(x,y)$ of the illustrative case where $f_1=0.6576, f_2=0.0835$.}
	\label{fig-potential}
\end{figure}

We consider a two-dimensional periodic medium with square elementary cells $[0,\eta]^2$ of side-length $\eta \ll 1$.  The solution $T_\eta$ of the steady-state heat equation with fast oscillating parameters $A_\eta=k(x/\eta,y/\eta)$
\begin{equation}
\centering
\nabla \cdot \left( k_\eta \nabla T_\eta\right)=0
\end{equation}
two-scale converges \cite{allaire1992homogenization}, when $\eta$ tends to zero, to the solution $T_{hom}$ of the homogenized heat equation  
\begin{equation}
\centering
\nabla \cdot \left( [k_{hom}] \nabla T_{hom}\right)=0.
\end{equation}
The effective property of the periodic medium is given by
\begin{equation}
\centering
[k_{hom}]=\left(\begin{array}{cc}
\langle k \rangle - \langle k \partial_x V_1 \rangle & -\langle k \partial_x V_2 \rangle\\
-\langle k \partial_y V_1 \rangle & \langle k \rangle - \langle k \partial_y V_2 \rangle\\
\end{array}\right) \label{eqhoma},
\end{equation}\
\noindent where $\partial_x:=\partial/\partial x$, $\partial_y:=\partial/\partial y$ and $<.>$ denotes the mean operator over the periodic cell. $V_1(x,y)$ and $V_2(x,y)$ are solutions defined up to an additive constant of auxiliary problems of thermostatic type on the periodic cell \cite{zolla2003artificial}:
\begin{equation}\label{GrindEQ__2.3_}
\centering
\left\lbrace
\begin{array}{c}
\nabla \cdot \left[ k(x,y) \nabla (V_1-x)\right]=0 \\
\nabla \cdot \left[ k(x,y) \nabla (V_2-y)\right]=0
\end{array}.
\right.
\end{equation}
We solve the auxiliary problems in weak form by using COMSOL Multiphysics, which sets up the finite element problem with periodic conditions imposed to the field on opposite ends of the elementary cell. We note in passing $V_1$ and $V_2$ are unique solutions of Eq. (\ref{GrindEQ__2.3_}) up to additive constants, but these constants do not affect the homogenized conductivity, as one can see that in Eq. (\ref{eqhoma}) only the partial derivatives of $V_1$ and $V_2$ are involved. The potentials $V_1(x,y)$ and $V_2(x,y)$ of an illustrative case ($f_1=0.6576, f_2=0.0835$) are shown in Fig. \ref{fig-potential}, where we obtain the effective conductivity as

\begin{equation}
\centering
[k_{hom}]=\begin{pmatrix}
66.01 & 3.3 \; 10^{-11}\\
-1.2 \; 10^{-11} & 263.97
\end{pmatrix} \label{potential}
=\begin{pmatrix}
  k_1 & 0 \\ 0 & k_2
 \end{pmatrix}
.
\end{equation}\
It is noticed that in Eq. (\ref{potential}) the off-diagonal components are almost negligible and originate from numerical errors. The resulting spurious artificial anisotropy can be safely ignored. We emphasize that with the finite element method, the effective tensor can be obtained for any periodic composite and that the same technique was implemented before in acoustics \cite{pomot2020acoustic} and electromagnetism \cite{zolla2003artificial}, in which case the effective density and permittivity tensors can be deduced from the same annex problems as in our thermal case.

The effective properties of the medium can be tuned by several design parameters such as geometry (here we use area fractions $f_1$ and $f_2$) and material properties (thermal conductivities $k_A$ and $k_B$). We restrict our attention here to predefined material properties and set the geometrical parameters as variables. Our goal is to find the set of geometrical parameters which properly mimic the homogenized medium. Therefore, the geometry of elementary cells should be tuned to obtain desired equivalent properties. Considering the heavy workload of the trial-and-error method, we implement an Optimal Latin hypercube technique to solve the problem. We note here that the two-scale homogenization technique has been already used to design a thermal concentrator similar to the one shown in Fig. \ref{fig-illustration} and to show that a concentrator consisting of concentric layers would require some complex valued conductivities with sign-shifting imaginary parts \cite{petiteau2015thermal}. In the present case, we investigate doubly periodic designs and thus we do not face such pitfalls.

\subsection{Optimal Latin Hypercube Design}

Optimal Latin Hypercube Sampling (OLHS) technique is applied to optimize the spatial positions of control points. The aim of this process is to design a matrix where the sample points spread as evenly as possible within the design region. This method is efficient and robust due to its enhanced stochastic evolutionary algorithm and significant reduction in matrix calculations to evaluate new/modified designs during searching \cite{park1994optimal}. In this work, we pre-define the two materials as air ($k_A=0.026$) and copper ($ k_B=400$) in units of $\rm{W m^{-1} K^{-1}}$ and select the geometrical (area fraction) parameters $f_1$ and $f_2$ as design variables. The ranges of these variables are defined as $0.5<f_1<0.8$ and $0.05<f_2<0.1$ after initial estimation. We generate one hundred sample points by OLHS technique, and calculate corresponding effective thermal conductivity of the periodic medium for each sample point using the two-scale homogenization theory. Details of the generated sample points and calculated results are listed in Table S1. We emphasize the importance of choosing proper ranges of the variables. If, for instance, we use much larger ranges for $f_1$ and $f_2$, the derived values can be less accurate unless a more refined discretization (more sample points) is chosen.

\begin{figure*}[!tbh]
	\centering
	\includegraphics[width=18cm]{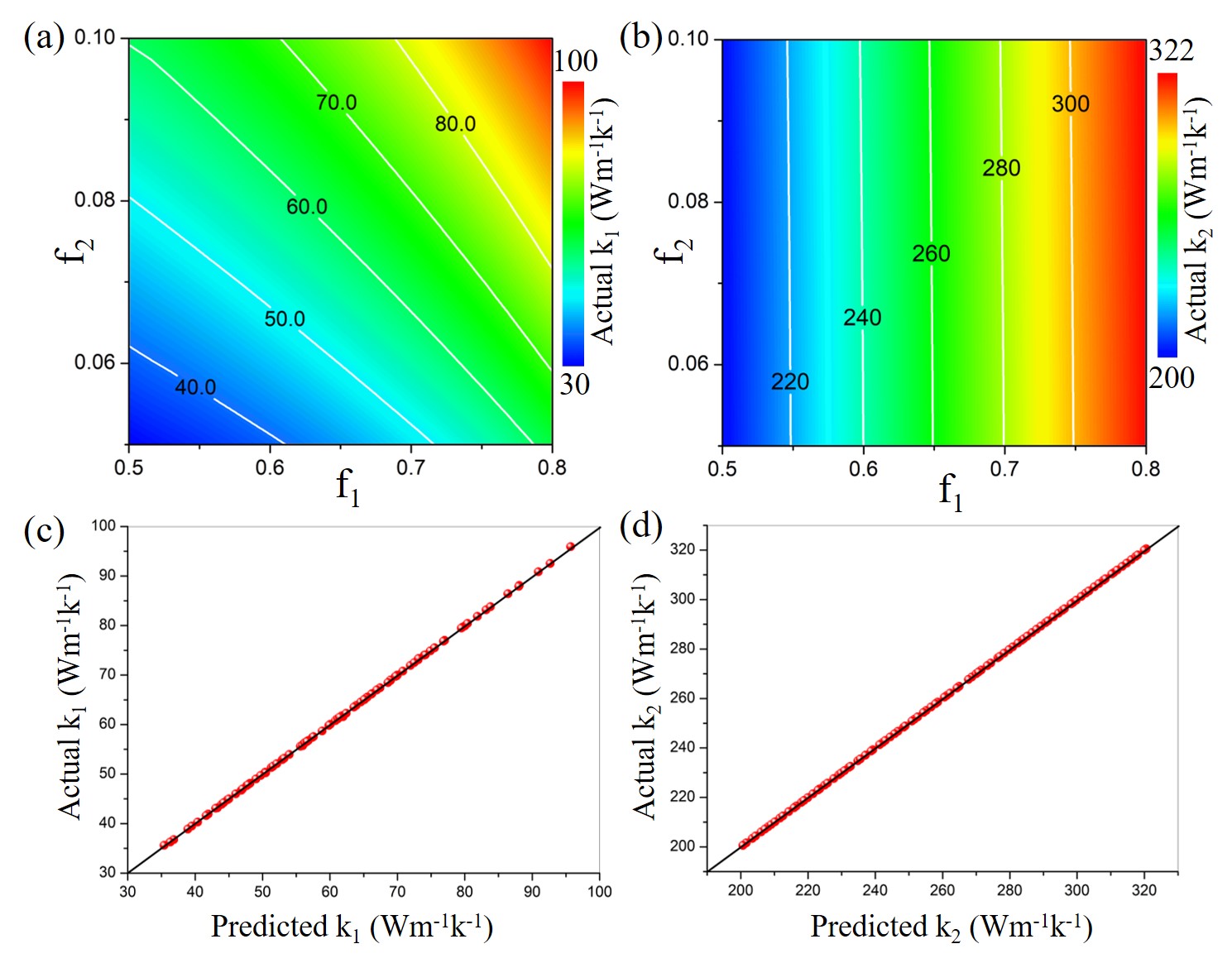}
	\caption{Panels (a) and (b) show the relation between design targets (heat conductivities $k_1$ and $k_2$) and design variables (geometrical parameters $f_1$ and $f_2$). The results are obtained basing on the chosen sample points and using two-scale homogenization theory. Panels (c) and (d) show good agreement between actual values and predicted values of the heat conductivities, demonstrating the reliability of the surrogate model. The indices $k_1$ and $k_2$ denote heat conductivity of the elementary cell in the x and y directions, respectively.}	
	\label{fig-relation}
\end{figure*}

We then create an approximation surrogate model from the obtained one hundred space samples \cite{mak2000estimation}. The surrogate model is built by the Elliptical Basis Function Neural Network technique which establishes a relation between design targets ($k_1$ and $k_2$) and variables ($f_1$ and $f_2$), as shown in Fig. \ref{fig-relation}.  

We use two estimators to evaluate the reliability of the surrogate model, the coefficient of determination ($R^{2}$) and the root mean square error (RMSE). These are defined as
\begin{align}
& R^{2} = 1-\frac{\sum_{i=1}^{m}(y_{i}-\hat{y_{i}})^2}{\sum_{i=1}^{m}(y_{i}-\bar{y_{i}})^2},  \\
& \mathrm{RMSE} = \sqrt{\frac{\sum_{i=1}^{m}(y_{i}-\hat{y_{i}})^2}{m}},
\end{align}
\noindent where $y_{i}$ and $\hat{y_{i}}$ are respectively the real value and the predicted value of the objective function over the same sample points, $\bar{y_{i}}$ is the mean value of all objective functions and $m$ is the total number of sampling points.
The closer $R^{2}$ is to 1 and RMSE is to 0, the more accurate the model. In Fig. \ref{fig-relation} (c-d), we can observe that the predicted values are in good agreement with the actual values. 

The calculated $R^{2}$ and RMSE are listed in Table \ref{tab1}. The average error and the maximum error among all samples are also shown. In the constructed surrogate models, $R^{2}$ is larger than 0.99997 and RMSE are smaller than 0.00012. The maximum error remains close to 0, which demonstrates that the surrogate models are accurate. 

\begin{table}[!tbh]
	\centering
	\captionsetup{singlelinecheck=off, skip=6pt}
	\caption{\newline Accuracy of the constructed surrogate model.}
	\begin{tabular*}{6.5cm}{ccc}
		\hline
		Error type & $k_1$ & $k_2$ \\
		\hline
		RMSE  & $1.2 \; 10^{-3}$ & $4.88 \; 10^{-5}$ \\
		Average  & $6.45 \; 10^{-4}$ & $3.52 \; 10^{-5}$  \\
		Maximum   & $5.47 \; 10^{-3}$ & $1.88 \; 10^{-4}$ \\
		$R^{2}$   & $0.9997$ & $1$ \\
		\hline
	\end{tabular*}
	\label{tab1}
\end{table}

Now we proceed with the inverse homogenization problem, that is, we want to find the particular set of design parameters which best mimic the transformed medium \cite{cherkaev2001inverse}. The optimization problem amounts to minimizing the objective function: 
\begin{equation}\label{GrindEQ__2.4_}
\centering
E(k_1,k_2,\alpha, \beta)= \alpha \left| k_1-k_r \right| + \beta \left| k_2-k_\theta \right|,
\end{equation}
\noindent where $\alpha$ and $\beta$ are weighting factors for the two objective sub-functions such that $\alpha+\beta=1$. The function $E(k_1,k_2,\alpha, \beta)$ measures the overall difference between obtained and objective values. We define $\alpha=\beta=0.5$, i.e. equal weights for the diagonal tensor elements $k_1$ and $k_2$. The principle is to obtain a global minimum within the discrete solution space. We solve this inverse problem by using a Non-Dominated Sorting Genetic Algorithm approach \cite{deb2002fast}. The one hundred random structures, corresponding to the parameter space samples obtained by the OLHS method, form the first generation. During the search process, the population size and the number of generations are defined as 12 and 200, respectively. New generations are created using crossover and mutation processes. We set the mutation distribution index and crossover distribution index as 20 and 10, respectively. The crossover probability is set as 0.9. For the sake of clarity, we do not detail the well-established genetic algorithm approach. In short, the Non-Dominated Sorting Genetic Algorithm performs well enough in our case and enables us to determine efficiently the desired design parameters.

Following this approach, we obtain the desired set of parameters as $f_1=0.6576$ and $f_2=0.0835$ for $C=2$. We then implement the microstructure and calculate the corresponding heat conductivities. The result are listed in Table.\ref{tab2}. It can be seen that the obtained parameters closely mimic the desired transformed medium.

\begin{table}[!tbh]
	\centering
	\captionsetup{singlelinecheck=off, skip=6pt}
	\caption{\newline Comparison of predicted value and targeted value for the derived set of parameters.} 
	\begin{tabular*}{9cm}{cccc}
		\hline
		& Predicted & Targeted & \tabincell{c}{Relative \\ difference} \\
		\hline
		$k_1 (\rm{W m^{-1} K^{-1}})$  & 263.97 & 264 & 0.01\% \\
		$k_2 (\rm{W m^{-1} K^{-1}})$  & 66.01 & 66 & 0.02\% \\
		\hline
	\end{tabular*}
	\label{tab2}
\end{table}

\section{Scheme validation and discussion}
\subsection{Recipe for experimental realization}
We now turn to the theoretical recipe for realizing thermal harvesting based on optimized composite microstructures. Note that the heat conductivity in Eq. \eqref{GrindEQ__2.2_} is expressed in $(r,\theta)$ polar coordinates while the microstructure is designed in $(x,y)$ Cartesian coordinates. To build a cylindrical thermal harvesting device, we discretize the shell region into numerous units and transplant microstructures with matched geometrical parameters into each unit. We design in this paper a discrete thermal harvesting device with 15 radial layers and 45 tangential sectors. Similarly to what has been done in Ref. \cite{pomot2020acoustic}, we could have considered an increasing number of layers and sectors to very accurately approximate the idealized thermal concentrator parameters. However, we focus here on a practically implementable design. The materials constituting the microstructure are naturally available materials: copper (material A) and air (material B). It is understood that the device could be built from other materials. We use copper and air here considering their applicability for realistic experiments.

\begin{figure}[!tbh]
	\centering
	\includegraphics[width=9cm]{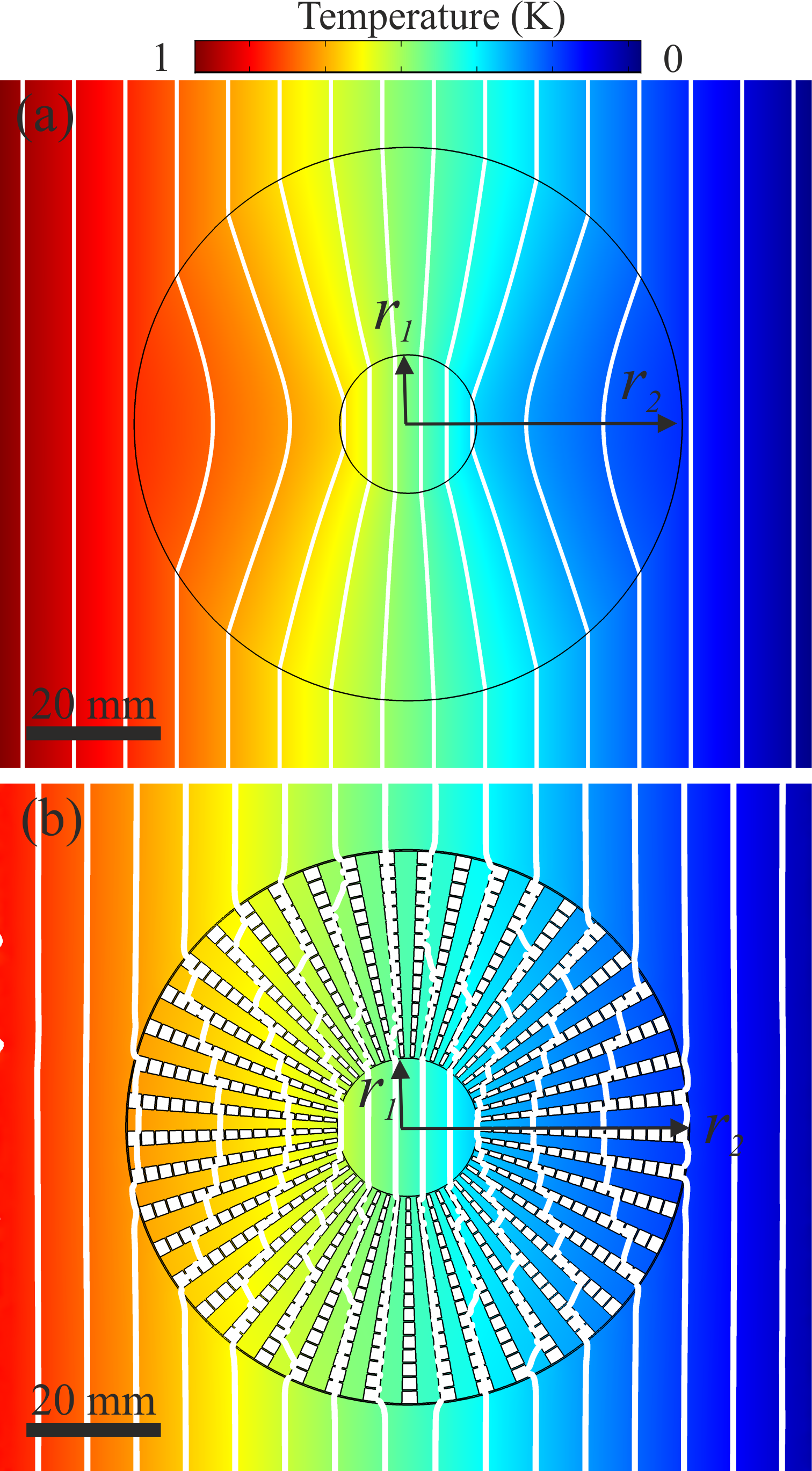}
	\caption{Temperature fields in the steady state for (a) the ideal case and (b) the proposed thermal harvesting device. In both cases, heat flows are concentrated to the inner domain without much perturbations to the external thermal field.}
	\label{fig-real}
\end{figure}

Numerical calculations were conducted, where temperatures at the left and right boundary are respectively imposed as $1 \rm{~K}$ and $0 \rm{~K}$, for easiness in normalization. We apply Neumann (perfect insulator) conditions at other boundaries. As indicated in the scheme (Fig. \ref{fig-real}), iso-thermal lines are significantly compressed to the inner domain ($r\leq r_1$). Hence, heat flux density in the inner domain is enlarged, implying that more heat energy is concentrated into the central region. In addition, iso-thermal lines in the background are uniform with little perturbations. That is, the heat energy is harvested and concentrated into the central region without much perturbation of the external thermal field. We emphasize that similar computations would also hold for time-harmonic acoustic and electromagnetic equations. 

We further conduct a quantitative analysis of the thermal harvesting behavior of the proposed scheme. Two measurement lines are defined. An horizontal line ($y=0$) is selected to reveal perturbations of the external thermal field whereas a vertical line ($x=r_2$) is chosen to illustrate the heat harvesting efficiency. We also build an additional contrast plate that occupies the same area as the harvesting device but that is composed only of a homogeneous background medium. The following index $\eta$ is defined to evaluate the energy harvesting efficiency as
\begin{equation} \label{GrindEQ__3.1_}
\centering
\eta=\left| \frac{T_{x=r_1}-T_{x=-r_1}}{T_{x=r_2}-T_{x=-r_2}} \right|.
\end{equation}
\noindent An index $M_{V}$ is introduced to characterize the perturbations to external fields (\textit{i.e.} the thermal neutrality of the concentrator)
\begin{equation} \label{GrindEQ__11_} 
\centering
M_{V} =\frac{\int _{\Omega }\left|T(x,y,z)-T_{r}(x,y,z) \right|d\Omega  }{\int _{\Omega }d\Omega  },                                                                      
\end{equation} 
\noindent where $\Omega $ denotes the probe domain of external thermal fields and $T_r$ represents the temperature field of the homogeneous medium. The index $M_{V}$ reveals all perturbations to the external heat profile \cite{ji2019thermal}.

\begin{figure*}[!tbh]
	\centering
	\includegraphics[width=18cm]{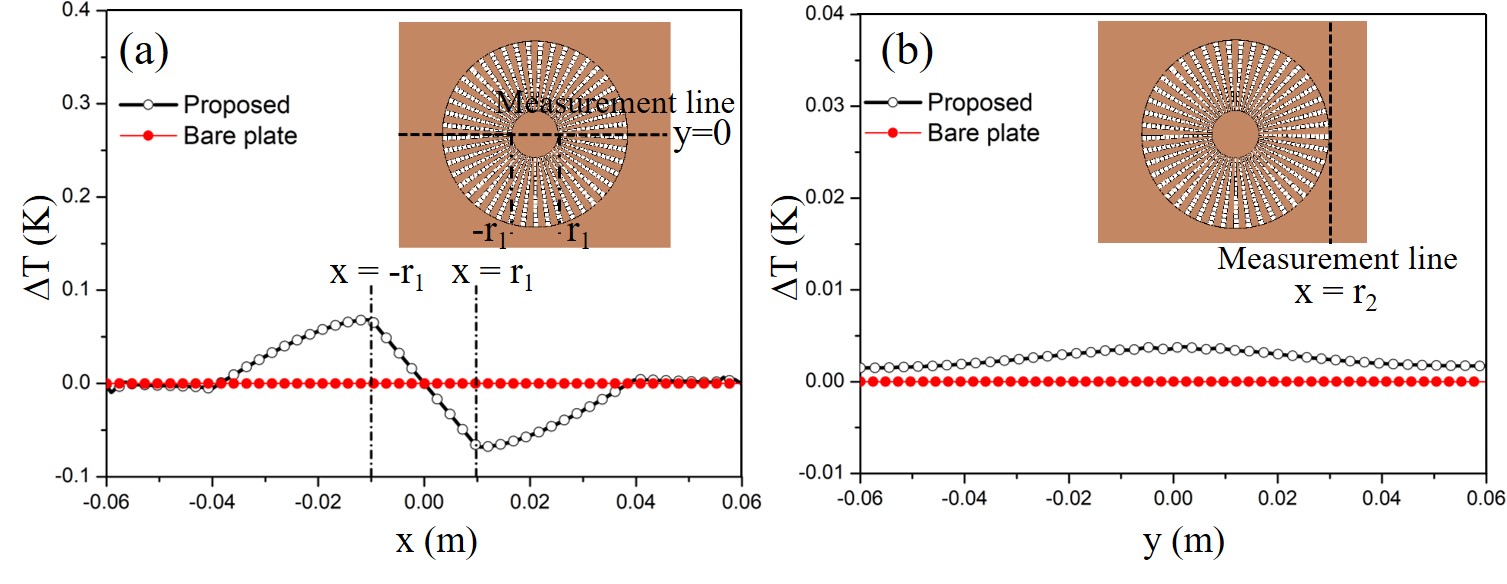}
	\caption{Comparison of temperature profiles between the harvesting device and the contrast plate along the measurement lines (a) $y=0$ and (b) $x=r_2$. All results are obtained in the steady-state and the insets denote positions of measurement lines. We define $\Delta T=\left|T(x,y,z)-T_{r}(x,y,z) \right|$ to quantitatively demonstrate the heat harvesting performances. Significant heat-focusing in the inner domain ($-r_1<x<r_1$) and minor fluctuations in external domain (along $x=r_2$) are both oberved, demonstrating good heat concentrating performances.}
	\label{fig-quant}
\end{figure*}

It is apparent in Fig. \ref{fig-quant}(a) that thermal energy is significantly concentrated into the inner domain, as a tight focusing with a local heat-intensity increase is observed. We list thermal gradients inside the inner domain (characterized by $\Delta T_{in}=\left|T_{x=r_1}-T_{x=-r_1} \right|$) and harvesting efficiencies for different cases in Tab. \ref{tab4}. The results indicate that thermal gradients in the central region are almost twice as large as in the contrast plate. The concentration efficiency is significantly lifted. Both results reach nearly theoretical values, demonstrating that the heat concentrating scheme is both effective and accurate.

\begin{table}[!tbh]
	\centering
	\captionsetup{singlelinecheck=off, skip=6pt}
	\caption{\newline Thermal harvesting performances for different cases.}
	\begin{tabular*}{8cm}{cccc}
		\hline
		Index & Ideal & Proposed & Bare plate \\
		\hline
		$\Delta T_{in} ~(\rm{K/m})$ & 0.4 & 0.39 & 0.2\\
		$\eta$ & 0.5 & 0.49 & 0.25\\
		$M_{V} ~(\rm{K})$ & 0 & 0.0025 & 0.0003\\
		\hline
	\end{tabular*}
	\label{tab4}
\end{table}

It can be observed in Fig. \ref{fig-quant}(b) that temperatures along the vertical measurement line are almost uniform, indicating that the external thermal field is only slightly influenced. We notice some minor perturbations in Tab. \ref{tab4} which are mainly due to the discretization process and to numerical errors. 

\begin{figure}[!tbh]
	\centering
	\includegraphics*[width=9cm, keepaspectratio=true]{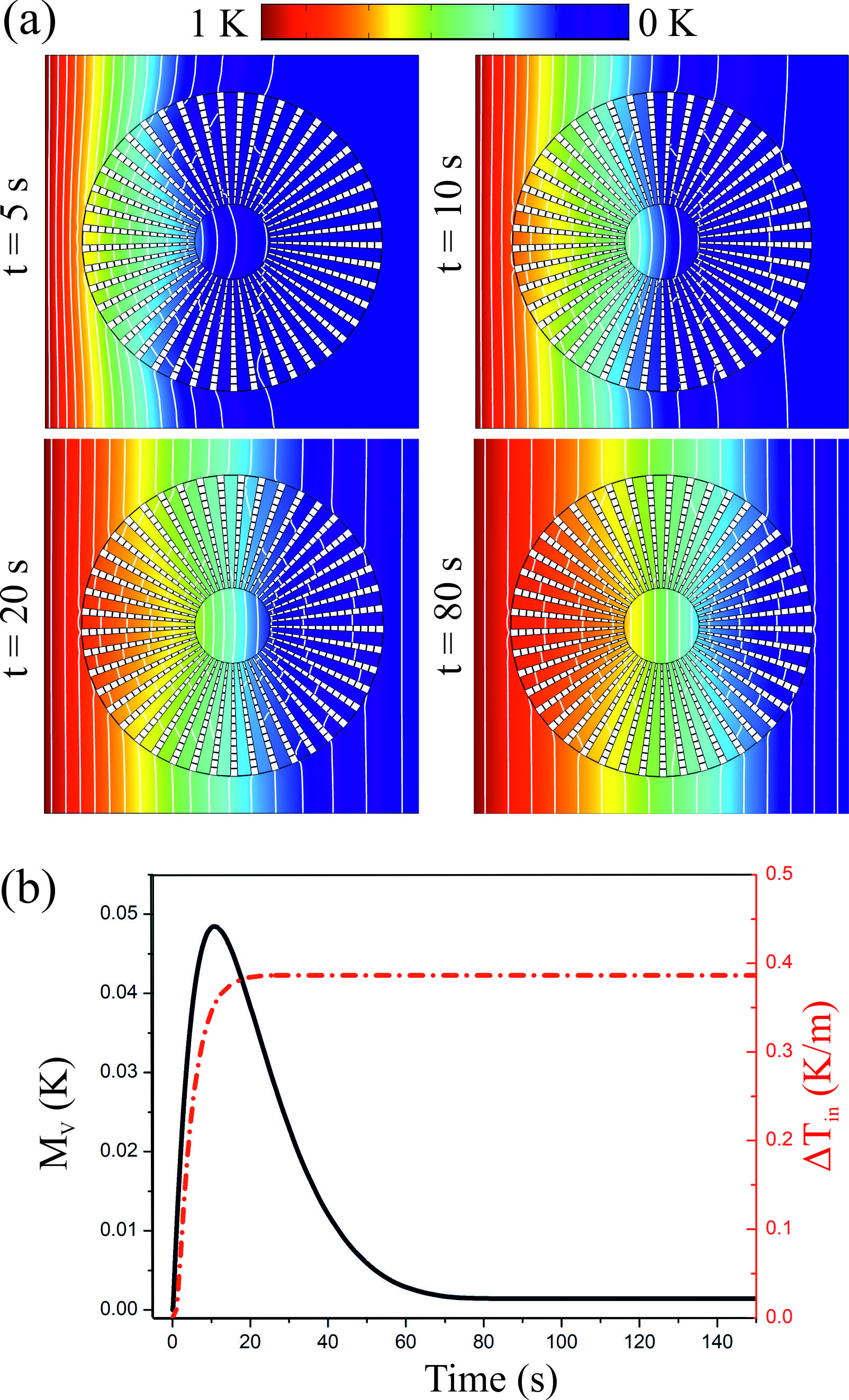}
	\caption{Thermal harvesting performances of the concentrator in the transient regime. (a) Temperature fields are illustrated at different times $t$. The performances get gradually better after a certain lapse of time. (b) Thermal gradients of the inner domain ($\Delta T_{in}=\left|T_{x=r_1}-T_{x=-r_1} \right|$) and perturbations to external thermal field ($M_V$) are plotted as functions of time. The quantitative results agree well with those revealed in (a).}
	\label{fig-dynamic}
\end{figure}

Thus far, all that has been achieved for heat can be directly translated to airborne acoustics (with rigid inclusions) and electromagnetism (with perfectly conducting inclusions, in the case when the magnetic field is polarized perpendicular to the xy-plane), since governing equations are identical in the static limit, and so results in Fig. 6 and 7 hold for the acoustic and electromagnetic counterparts of the thermal concentrator. We would like to investigate now the diffusive nature of heat conduction. We define same boundary conditions as the steady state case and focus here on the evolution of thermal harvesting performance over time. Temperature distributions at different time steps $t$ are shown in Fig. (\ref{fig-dynamic}), where thermal gradients and perturbations are also plotted as a function of time. It is observed that perturbations of the external thermal profile increase and then decrease after a certain lapse of time, whereas the thermal gradient of the object increases gradually toward its maximum value.
Generally, the proposed design performs well for harvesting thermal energy, with significant heat flux concentrated in the inner domain and little perturbation to the outside thermal field, once the permanent regime has been reached. We stress that the design process proves feasible as the created device works efficiently and converges to almost the same harvesting performance as in the ideal case. 

\begin{figure}[!tbh]
	\centering
	\includegraphics*[width=9cm, keepaspectratio=true]{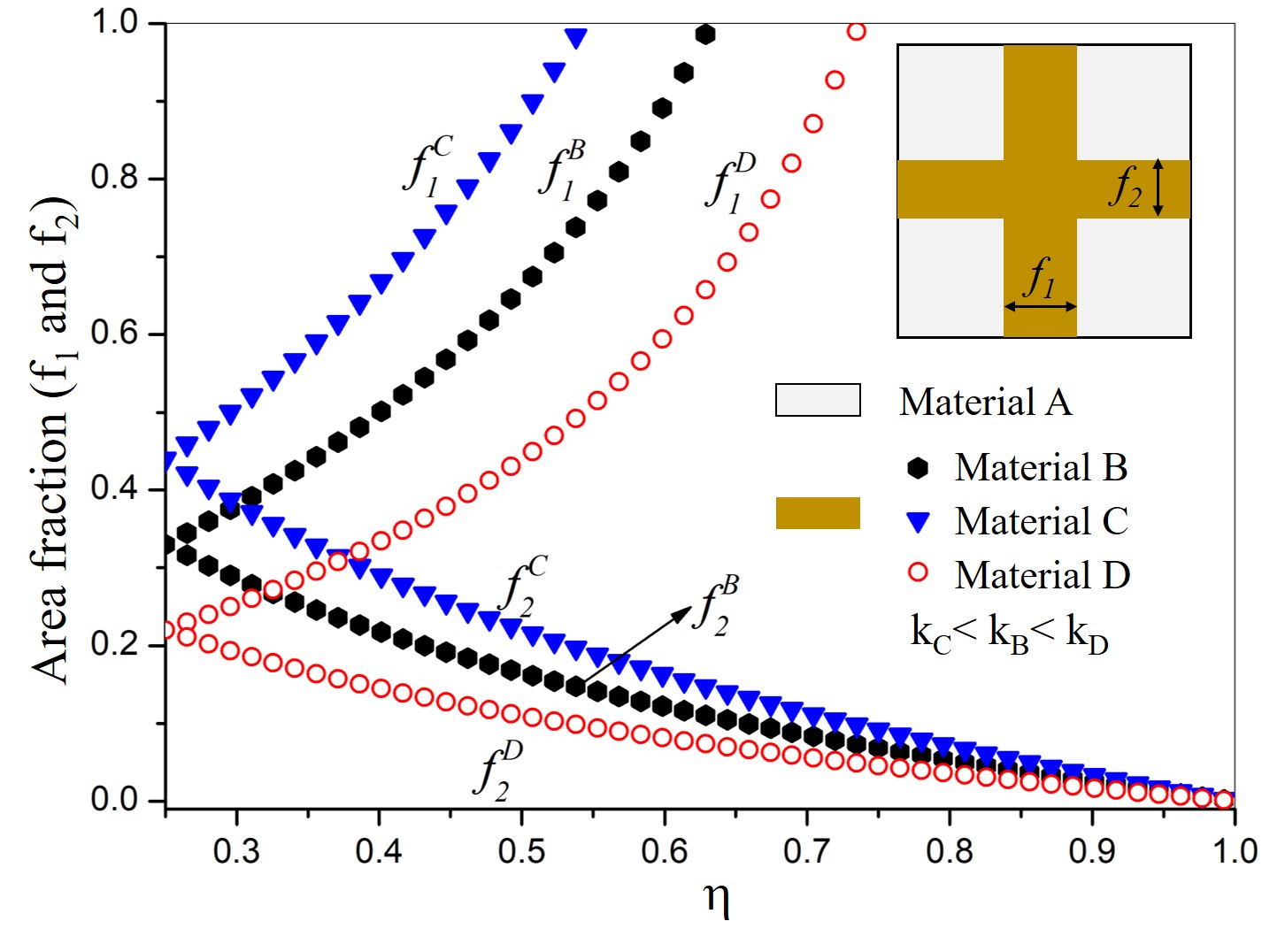}
	\caption{Relation between design variables (area fractions $f_1$ and $f_2$) and the concentration efficiency. The inset shows the studied elementary cell where one material is pre-defined as $k_A=0$ and another material is varied. In addition to material B ($k_B=400 \rm{~W m^{-1} K^{-1}}$), we also considered material C ($k_C=300 \rm{~W m^{-1} K^{-1}}$) and material D ($k_D=600 \rm{~W m^{-1} K^{-1}}$) for a comparison.}
	\label{fig2-parameter}
\end{figure}

The heat concentration efficiency is directly determined by the constant $C$ \cite{han2013theoretical}, which in turn can be approached by the material properties and the geometry of the microstructures. To gain more insights into the underlying mechanism of the microstructure, we derive and show in Fig. \ref{fig2-parameter} the relation between design variables (area fraction $f_1$ and $f_2$) and the heat concentration efficiency. For comparison, we further assume two other materials C and D to substitute for material B, with heat conductivity $k_C<k_B<k_D$. It is observed that higher concentration efficiency requires larger $f_1$ and smaller $f_2$ which directly indicates larger geometrical anisotropy. Besides, we notice that smaller design variables are needed to achieve a given efficiency for a larger conductivity of the second material (for constant material A).  Larger material anisotropy also allows for a larger maximum concentration efficiency.  

\subsection{Additional microstructures}

\begin{figure}[!tbh]
	\centering
	\includegraphics*[width=9cm, keepaspectratio=true]{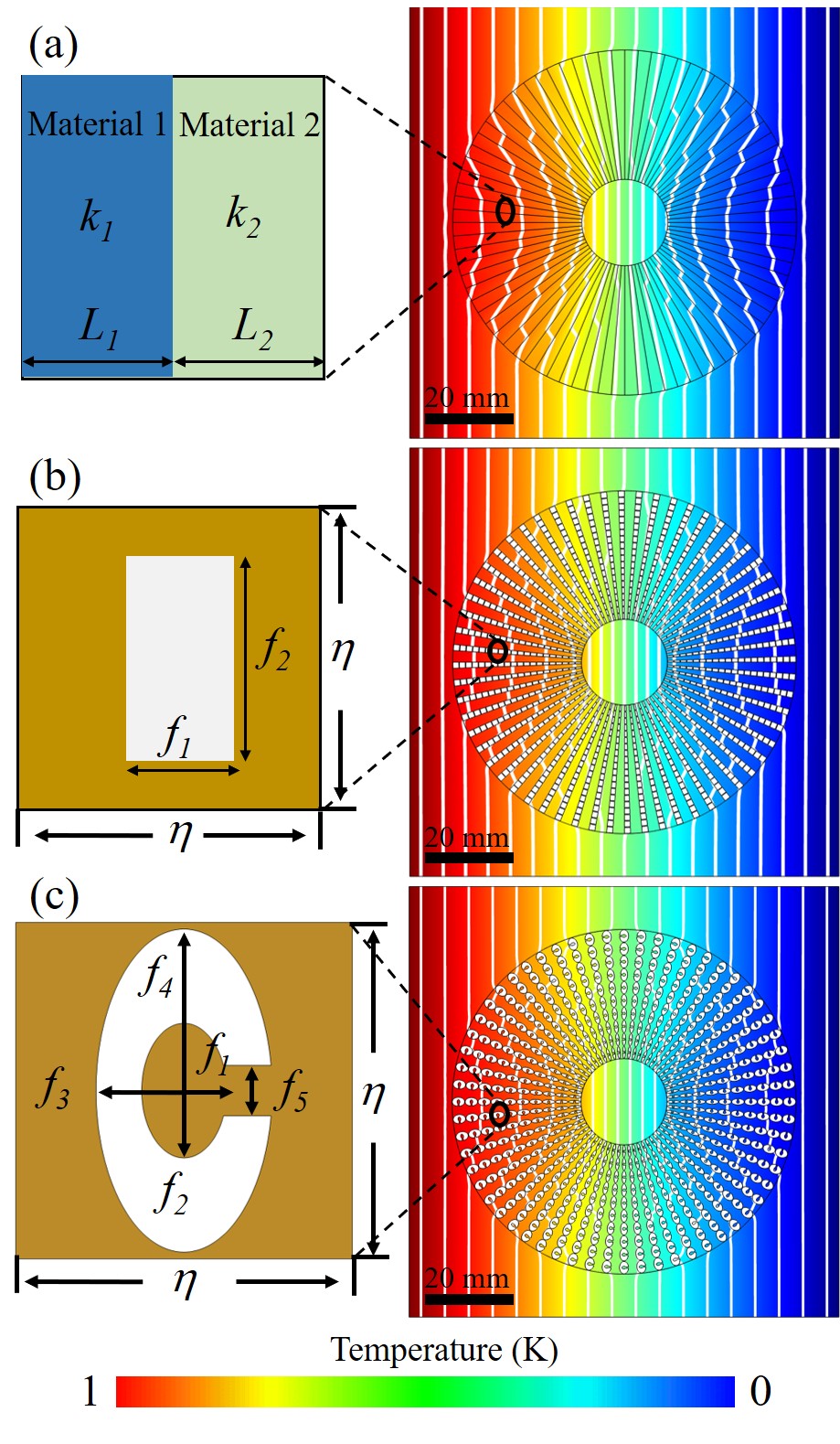}
	\caption{Additional optimized microstructures and corresponding thermal harvesting performance. (a) Laminar structures defined by two material parameters ($k_1$ and $k_2$) if $L_1=L_2$ and two additional geometrical parameters if $L_1 \neq L_2$. The optimized parameters are $k_1=0.26782$ and $k_2=3.7328$ if we impose $L_1=L_2$. (b) Elementary cell defined by two geometrical parameters $f_1=0.34238, f_2=0.91675$. (c) Split ring structure defined by five geometrical parameters. The optimized parameters are $f_1=0.1037$, $f_2=0.1261$, $f_3=0.4799$, $f_4=0.2614$, and $f_5=0.0994$. Following the proposed design process, we can deal with different microstructures and obtain desired thermal harvesting performances.}
	\label{fig2-additional}
\end{figure}

Note that the microstructure we employed in Fig. \ref{fig-cell} is not the unique choice. One can use other elementary cells as long as they are appropriate to obtain the required anisotropy. To validate the wide-range application of our approach, we further present several other elementary cells and build corresponding devices. Steady state simulations are conducted with same boundaries with the aforementioned case. In Fig. \ref{fig2-additional}(a), we set material properties as design variables and recover the widely used solar-shaped device. In Fig. \ref{fig2-additional}(b), we obtain the same design as in Fig. \ref{fig-illustration} but based on a different elementary cell obtained by a sub-lattice translation. 
Besides, we show in Fig. \ref{fig2-additional}(c) a split ring element cell for which it seems unlikely that the effective medium theory could be easily applied. Good harvesting performance is again achieved without much perturbations to the external field. As a note, the symmetry of the split ring element cell is reduced compared to the other cases, but the effective tensor remains diagonal. The strength of our approach becomes apparent when dealing with such more complex geometries, where the application of effective medium theory is far from trivial.

We stress that using the optimization approach in this paper high-performance thermal-harvesting designs can be obtained that meet given external constraints, such as maximum or minimum values for material properties and geometrical parameters.
We can search for the best solution among materials properties, microstructures and geometrical parameters within the full available range, providing a substantial flexibility and many degrees of freedom in practical applications.

We considered in this paper two-dimensional (2D) harvesting devices as fabrication and measurement are presumably simpler than in 3D. The proposed optimization method, however, is also well suited to the design of 3D thermal harvesting devices. A major difference is that more design variables are required in 3D, and thus more computational resources would be required to undertake such a study, for which the theoretical part of the present work would apply mutatis mutandis.

\section{Conclusion}

We proposed a general method to design thermal-energy harvesting devices from naturally available materials. We designed composite microstructures and calculated their effective conductivity by the two-scale homogenization technique. We then implemented the Optimal Latin hypercube method to obtain design parameters that can best mimic the transformed medium. We built a harvesting device model based on the obtained optimized microstructures and demonstrated good thermal harvesting performance, thereby validating the effectiveness of our design method. 

The optimization method adds great flexibility to the constraints that can be imposed on devices, which allows one to find the best possible solutions with different geometrical structures and component materials. We stress that the flexibility and simplicity of the method is a good addition to existing heat manipulation techniques, including other thermal functionalities, \textit{i.e.} cloaking and illusion. It also paves a path for novel optical, acoustic and electromagnetic devices based on form-invariant governing equations.

\bigskip
\section*{Acknowledgments} 
This work was supported by the EIPHI Graduate School [grant number ANR-17-EURE-0002]; the French Investissements d'Avenir program, project ISITEBFC [grant number ANR-15-IDEX-03]; and the National Natural Science Foundation of China [grant numbers 11732002 and 11672089].
\label{}






\begin{thebibliography}{10}
\expandafter\ifx\csname url\endcsname\relax
  \def\url#1{\texttt{#1}}\fi
\expandafter\ifx\csname urlprefix\endcsname\relax\def\urlprefix{URL }\fi
\expandafter\ifx\csname href\endcsname\relax
  \def\href#1#2{#2} \def\path#1{#1}\fi

\bibitem{pendry2006controlling}
J.~B. Pendry, D.~Schurig, D.~R. Smith, Controlling electromagnetic fields,
  Science 312~(5781) (2006) 1780--1782.

\bibitem{kadic2013metamaterials}
M.~Kadic, T.~B{\"u}ckmann, R.~Schittny, M.~Wegener, Metamaterials beyond
  electromagnetism, Reports on Progress in Physics 76~(12) (2013) 126501.

\bibitem{milton2006cloaking}
G.~W. Milton, M.~Briane, J.~R. Willis, On cloaking for elasticity and physical
  equations with a transformation invariant form, New Journal of Physics 8~(10)
  (2006) 248.

\bibitem{buckmann2014elasto}
T.~B{\"u}ckmann, M.~Thiel, M.~Kadic, R.~Schittny, M.~Wegener, An
  elasto-mechanical unfeelability cloak made of pentamode metamaterials, Nature
  Communications 5 (2014) 4130.

\bibitem{achaoui2020cloaking}
Y.~Achaoui, A.~Diatta, M.~Kadic, S.~Guenneau, Cloaking in-plane elastic waves
  with swiss rolls, Materials 13~(2) (2020) 449.

\bibitem{kadic2020elastodynamic}
M.~Kadic, M.~Wegener, A.~Nicolet, F.~Zolla, S.~Guenneau, A.~Diatta,
  Elastodynamic behavior of mechanical cloaks designed by direct lattice
  transformations, Wave Motion 92 (2020) 102419.

\bibitem{schittny2014invisibility}
R.~Schittny, M.~Kadic, T.~B{\"u}ckmann, M.~Wegener, Invisibility cloaking in a
  diffusive light scattering medium, Science 345~(6195) (2014) 427--429.

\bibitem{schurig2006metamaterial}
D.~Schurig, J.~Mock, B.~Justice, S.~A. Cummer, J.~B. Pendry, A.~Starr, D.~R.
  Smith, Metamaterial electromagnetic cloak at microwave frequencies, Science
  314~(5801) (2006) 977--980.

\bibitem{cummer2007one}
S.~A. Cummer, D.~Schurig, One path to acoustic cloaking, New Journal of Physics
  9~(3) (2007) 45.

\bibitem{guenneau2012transformation}
S.~Guenneau, C.~Amra, D.~Veynante, Transformation thermodynamics: cloaking and
  concentrating heat flux, Optics Express 20~(7) (2012) 8207--8218.

\bibitem{farhat2009ultrabroadband}
M.~Farhat, S.~Guenneau, S.~Enoch, Ultrabroadband elastic cloaking in thin
  plates, Physical Review Letters 103~(2) (2009) 024301.

\bibitem{li2008hiding}
J.~Li, J.~B. Pendry, Hiding under the carpet: a new strategy for cloaking,
  Physical Review Letters 101~(20) (2008) 203901.

\bibitem{yang2015invisible}
T.~Yang, X.~Bai, D.~Gao, L.~Wu, B.~Li, J.~T. Thong, C.-W. Qiu, Invisible
  sensors: Simultaneous sensing and camouflaging in multiphysical fields,
  Advanced Materials 27~(47) (2015) 7752--7758.

\bibitem{liu2012dc}
M.~Liu, Z.~Lei~Mei, X.~Ma, T.~J. Cui, Dc illusion and its experimental
  verification, Applied Physics Letters 101~(5) (2012) 051905.

\bibitem{hu2018illusion}
R.~Hu, S.~Zhou, Y.~Li, D.-Y. Lei, X.~Luo, C.-W. Qiu, Illusion thermotics,
  Advanced Materials 30~(22) (2018) 1707237.

\bibitem{liu2007far}
Z.~Liu, H.~Lee, Y.~Xiong, C.~Sun, X.~Zhang, Far-field optical hyperlens
  magnifying sub-diffraction-limited objects, Science 315~(5819) (2007)
  1686--1686.

\bibitem{li2009experimental}
J.~Li, L.~Fok, X.~Yin, G.~Bartal, X.~Zhang, Experimental demonstration of an
  acoustic magnifying hyperlens, Nature Materials 8~(12) (2009) 931--934.

\bibitem{kadic2011plasmonic}
M.~Kadic, S.~Guenneau, S.~Enoch, S.~A. Ramakrishna, Plasmonic space folding:
  Focusing surface plasmons via negative refraction in complementary media, ACS
  nano 5~(9) (2011) 6819--6825.

\bibitem{freeman2017small}
J.~Freeman, I.~Guarracino, S.~A. Kalogirou, C.~N. Markides, A small-scale solar
  organic rankine cycle combined heat and power system with integrated thermal
  energy storage, Applied thermal engineering 127 (2017) 1543--1554.

\bibitem{amin2017thermal}
M.~Amin, N.~Putra, E.~A. Kosasih, E.~Prawiro, R.~A. Luanto, T.~Mahlia, Thermal
  properties of beeswax/graphene phase change material as energy storage for
  building applications, Applied Thermal Engineering 112 (2017) 273--280.

\bibitem{fan2008shaped}
C.~Fan, Y.~Gao, J.~Huang, Shaped graded materials with an apparent negative
  thermal conductivity, Applied Physics Letters 92~(25) (2008) 251907.

\bibitem{zeng2014experimental}
L.~Zeng, R.~Song, Experimental observation of heat transparency, Applied
  Physics Letters 104~(20) (2014) 201905.

\bibitem{schittny2013experiments}
R.~Schittny, M.~Kadic, S.~Guenneau, M.~Wegener, Experiments on transformation
  thermodynamics: molding the flow of heat, Physical Review Letters 110~(19)
  (2013) 195901.

\bibitem{han2014full}
T.~Han, X.~Bai, J.~T. Thong, B.~Li, C.-W. Qiu, Full control and manipulation of
  heat signatures: Cloaking, camouflage and thermal metamaterials, Advanced
  Materials 26~(11) (2014) 1731--1734.

\bibitem{peng2019three}
X.~Peng, R.~Hu, Three-dimensional illusion thermotics with separated thermal
  illusions, ES Energy \& Environment 6 (2019) 39--44.

\bibitem{huang2019macroscale}
S.~Huang, J.~Zhang, M.~Wang, R.~Hu, X.~Luo, Macroscale thermal diode-like black
  box with high transient rectification ratio, ES Energy \& Environment 6
  (2019) 51--6.

\bibitem{xu2018achieving}
G.~Xu, H.~Zhang, Y.~Jin, Achieving arbitrarily polygonal thermal harvesting
  devices with homogeneous parameters through linear mapping function, Energy
  Conversion and Management 165 (2018) 253--262.

\bibitem{chen2010transformation}
H.~Chen, C.~T. Chan, P.~Sheng, Transformation optics and metamaterials, Nature
  Materials 9~(5) (2010) 387.

\bibitem{narayana2012heat}
S.~Narayana, Y.~Sato, Heat flux manipulation with engineered thermal materials,
  Physical Review Letters 108~(21) (2012) 214303.

\bibitem{han2014experimental}
T.~Han, X.~Bai, D.~Gao, J.~T. Thong, B.~Li, C.-W. Qiu, Experimental
  demonstration of a bilayer thermal cloak, Physical Review Letters 112~(5)
  (2014) 054302.

\bibitem{zhang2020ellipsoidal}
X.~Zhang, X.~He, L.~Wu, Ellipsoidal bifunctional thermal-electric transparent
  device, Composite Structures 234 (2020) 111717.

\bibitem{xu2014ultrathin}
H.~Xu, X.~Shi, F.~Gao, H.~Sun, B.~Zhang, Ultrathin three-dimensional thermal
  cloak, Physical Review Letters 112~(5) (2014) 054301.

\bibitem{han2018full}
T.~Han, P.~Yang, Y.~Li, D.~Lei, B.~Li, K.~Hippalgaonkar, C.-W. Qiu,
  Full-parameter omnidirectional thermal metadevices of anisotropic geometry,
  Advanced Materials 30~(49) (2018) 1804019.

\bibitem{chen2015experimental}
F.~Chen, D.~Y. Lei, Experimental realization of extreme heat flux concentration
  with easy-to-make thermal metamaterials, Scientific Reports 5 (2015) 11552.

\bibitem{han2013theoretical}
T.~Han, J.~Zhao, T.~Yuan, D.~Y. Lei, B.~Li, C.-W. Qiu, Theoretical realization
  of an ultra-efficient thermal-energy harvesting cell made of natural
  materials, Energy \& Environmental Science 6~(12) (2013) 3537--3541.

\bibitem{mei2007effective}
J.~Mei, Z.~Liu, W.~Wen, P.~Sheng, Effective dynamic mass density of composites,
  Physical Review B 76~(13) (2007) 134205.

\bibitem{torrent2008acoustic}
D.~Torrent, J.~S{\'a}nchez-Dehesa, Acoustic cloaking in two dimensions: a
  feasible approach, New Journal of Physics 10~(6) (2008) 063015.

\bibitem{ji2018achieving}
Q.~Ji, G.~Fang, J.~Liang, Achieving thermal concentration based on fiber
  reinforced composite microstructures design, Journal of Physics D: Applied
  Physics 51~(31) (2018) 315304.

\bibitem{pomot2020acoustic}
L.~Pomot, C.~Payan, M.~Remillieux, S.~Guenneau, Acoustic cloaking: Geometric
  transform, homogenization and a genetic algorithm, Wave Motion 92 (2020)
  102413.

\bibitem{zolla2003artificial}
F.~Zolla, S.~Guenneau, Artificial ferro-magnetic anisotropy: homogenization of
  3d finite photonic crystals, in: IUTAM Symposium on Asymptotics,
  Singularities and Homogenisation in Problems of Mechanics, Springer, 2003,
  pp. 375--384.

\bibitem{allaire1992homogenization}
G.~Allaire, Homogenization and two-scale convergence, SIAM Journal on
  Mathematical Analysis 23~(6) (1992) 1482--1518.

\bibitem{petiteau2015thermal}
D.~Petiteau, S.~Guenneau, M.~Bellieud, M.~Zerrad, C.~Amra, Thermal concentrator
  homogenized with solar-shaped mantle, arXiv preprint arXiv:1508.05081 (2015).

\bibitem{park1994optimal}
J.-S. Park, Optimal latin-hypercube designs for computer experiments, Journal
  of Statistical Planning and Inference 39~(1) (1994) 95--111.

\bibitem{mak2000estimation}
M.-W. Mak, S.-Y. Kung, Estimation of elliptical basis function parameters by
  the em algorithm with application to speaker verification, IEEE Transactions
  on Neural Networks 11~(4) (2000) 961--969.

\bibitem{cherkaev2001inverse}
E.~Cherkaev, Inverse homogenization for evaluation of effective properties of a
  mixture, Inverse Problems 17~(4) (2001) 1203.

\bibitem{deb2002fast}
K.~Deb, A.~Pratap, S.~Agarwal, T.~Meyarivan, A fast and elitist multiobjective
  genetic algorithm: Nsga-ii, IEEE Transactions on Evolutionary Computation
  6~(2) (2002) 182--197.

\bibitem{ji2019thermal}
Q.~Ji, X.~Chen, G.~Fang, J.~Liang, X.~Yan, V.~Laude, M.~Kadic, Thermal cloaking
  of complex objects with the neutral inclusion and the coordinate
  transformation methods, AIP Advances 9~(4) (2019) 045029.

\end{thebibliography}
\end{document}